\documentclass[11pt]{article}
\usepackage{graphicx} 
\usepackage{blindtext}
\usepackage{float}
\usepackage{geometry}
\usepackage{amsmath}
\usepackage{caption}
\usepackage{amssymb}
\usepackage{subcaption}
\usepackage{hyperref}
\usepackage[numbers]{natbib}
\fontsize{12}{9}\selectfont
\begin{document}

\begin{titlepage}
\title{\textbf{Chaos in the Order of Finite Bernoulli Convolutions}\vspace{15mm}}

\author{ 
\textbf{Nadav Shrayer} \\ \href{mailto:nadavshraier@tauex.tau.ac.il}{nadavshraier@tauex.tau.ac.il}}

\date{
\emph{The Raymond and Beverly Sackler School of Physics and Astronomy}\\
\emph{Tel Aviv University, Ramat Aviv 69978, Tel Aviv, Israel}
 \\ [\baselineskip]
\today}
	
\maketitle
\vspace{15mm}

\begin{abstract}
In this note we explore numerically the finite Bernoulli convolutions. We show that with a suitable choice of parameter, it might serve as a toy model for intermittent energy cascade in fully developed turbulence. We then show how the crossings of $\beta$-expansions distribute in $\beta$, and suggest that it might highlight the parameters with enhanced overlap structure that are related to measures that are singular continuous. We later introduce a notion of order to the $\beta$-expansions based on the lexicographical order of the $N$-binary words, and observe that for most sampled adjacent pairs when $\beta=2$, the distance in their order increases exponentially when $\beta$ decreases from 2 to 1. This suggests 'chaotic' behavior, with the 'Lyapunov exponents' bunched into several clusters that depend on $N$. We end the note with some 'order plots' and an interesting connection between the finite $\beta$-compactum with $\beta=g$ (g being the golden ratio) and binary reflected Gray code.
\end{abstract}

\end{titlepage}

\section{Introduction}
The Bernoulli convolutions \cite{a12fbd66a1264f5ea55c6ec65fcc5fd0} arise from the probability measure of the sum 

\begin{equation}\label{BC sum}
    Y_{\lambda,N}=\sum_{n=0}^{N-1}\pm\lambda^n
\end{equation}

Here the $\pm1$ coefficients are random variables with a Bernoulli distribution (hence the name) of equal probability.
The measure is

\begin{equation}
\label{eq:finite BC}
\nu_{\lambda,N}=2^{-N}\sum_{\epsilon_0,\epsilon_1,\dots,\epsilon_{N-1}\in\{-1,1\}} \delta_{\sum^{N-1}_{k=0}\epsilon_k\lambda^k}
\end{equation}

and when $N\rightarrow\infty$, $\nu_{\lambda,N}$ weak-* converges to the infinite Bernoulli convolution $\nu_\lambda$, which also obeys the self similarity relation

\begin{equation}
\label{eq:self similarity}
\nu_\lambda=\frac{1}{2}(\nu_\lambda\circ T_{+}^{-1}+\nu_\lambda\circ T_{-}^{-1})
\end{equation}

that arises from the self-similar set of an IFS with the contractions $T_{\pm}(x)=\lambda x\pm1$.\par
The sums in the delta distributions in  (\ref{eq:finite BC}), over all polynomials of $\lambda$ with coefficients drawn from $\{-1,1\}$, hint at symbolic dynamics. In this picture we consider the set of all words of length $N$ made from binary digits $\mathcal{A}^N=\{0,1\}^N$, and the coding $\pi_\beta: \mathcal{A}^N\rightarrow \mathbb{R}$ given by the $\beta$-expansion

\begin{equation}
\label{eq:beta expansion}
\pi_\beta(\epsilon)=\sum_{k=0}^{N-1}\epsilon_k\beta^{-k}
\end{equation}

Where $\epsilon\in\mathcal{A}^N$ and $\beta=1/\lambda$. We denote the set of all $\beta$-expansions of $\epsilon\in\mathcal{A}^N$ as $\Omega_{\beta,N}$.\par
\begin{figure}[!ht]
    \centering
    \includegraphics[width=1\linewidth]{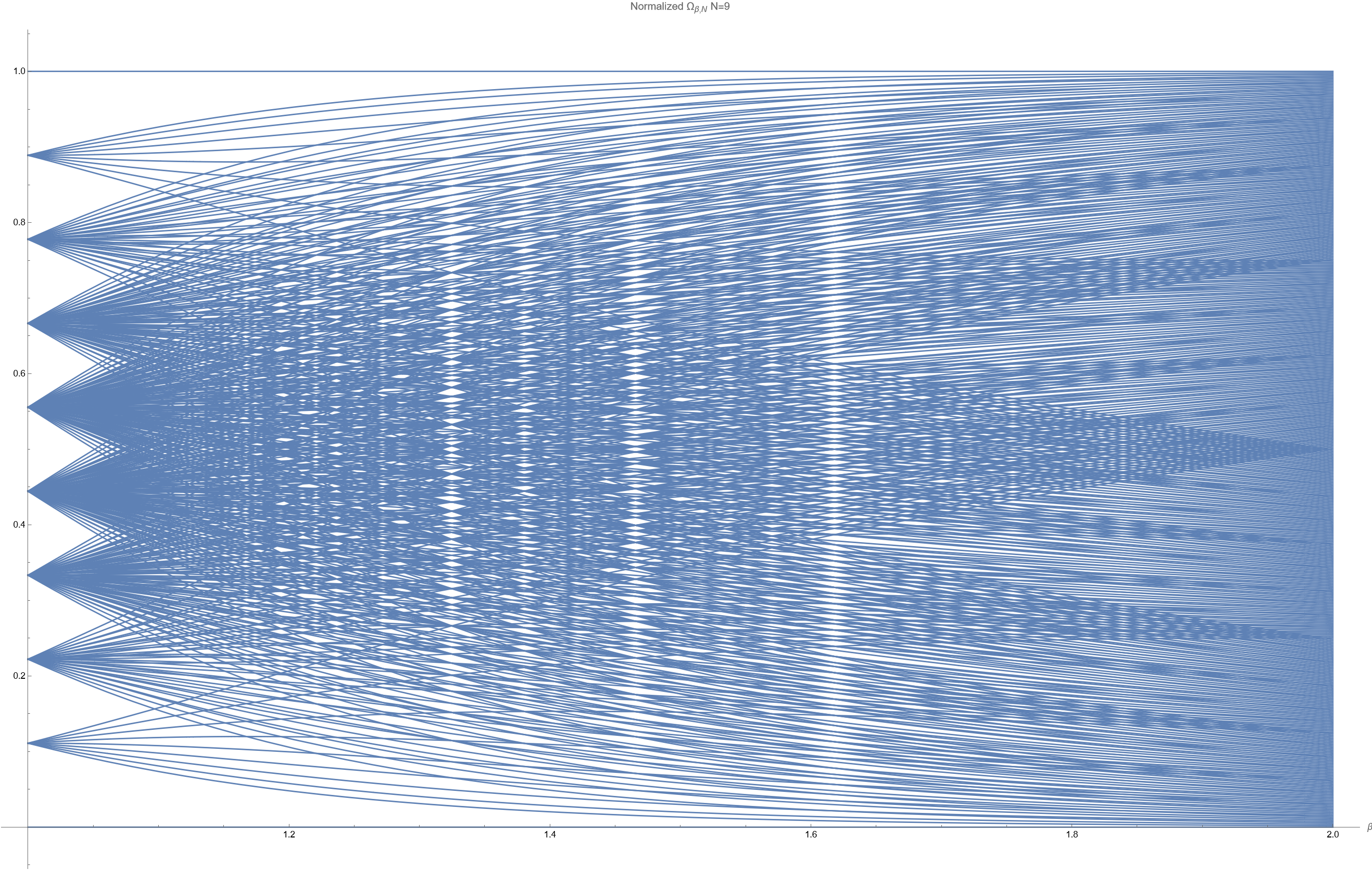}
    \caption{Plot of $\Omega_{\beta}$ normalized by the $\beta$-expansion of $\epsilon=\{1,1,\dots,1\}$, with $N=8$. Note that when $\beta=2$ all $\pi_\beta$ are equally spaced, and for $\beta=1$ $\Omega_{\beta}$ shows a Binomial distribution in $\pi_\beta$. For intermediate values of $\beta$, some values seem to form 'bright' columns that differ than the surroundings, which arise from values with relatively more crossings. The three most prominent seen in here correspond to the values $\beta_0\approx1.61803$, $\beta_1\approx1.46557$ and $\beta_2\approx1.32472$, which are all Pisot numbers. One can also spot "dark" columns, such as that of $\beta=\sqrt{2}$.} 
    \label{Normalized Polynomials n=9}
\end{figure}

 
We are interested in the structure of $\Omega_{\beta,N}$ and how it changes continuously with $\beta$, and we do so by looking at $\pi_\beta$ which are polynomials of $\beta$, as seen in Figure (\ref{Normalized Polynomials n=9}). 

Specifically, we are interested in the ordering of $\pi_\beta$ in $\Omega_{\beta,N}$ as $\beta$ changes. To keep track of the different $\pi_\beta$ we label them $\pi_\beta^n:=\pi_\beta(\epsilon_n)$, according to the lexicographical order of the sequences $\epsilon$ such that $\epsilon_n\prec_{lex}\epsilon_m$ when $n<m$. The order of $\pi_\beta^n\in\Omega_{\beta,N}$ is then defined as 

\begin{equation}
\label{eq:order of pi}
g(\pi^n_\beta):=2^{-N}|\{\pi_\beta^m|\pi_\beta^m<\pi^n_\beta\}|
\end{equation}

For $\beta\geq2$, this ordering is the same as the lexicographical order, such that if $\epsilon_n\prec_{lex}\epsilon_m$ then $g(\pi^n_\beta)<g(\pi^m_\beta)$. For $\beta<2$, there could be degeneracies in the values of $\pi_\beta$ when two or more sequences share the same $\beta$-expansion, which also carries out to a degeneracy in the ordering. These coincident $\pi_\beta$ can be seen in Figure (\ref{Normalized Polynomials n=9}) as crossings of the curves, and are the causes for changes in the ordering of $\Omega_{\beta,N}$.
From a numerical study of the crossings we get two main results: First, we calculate the distribution of crossings of a finite $\Omega_N$ in the interval $\beta\in(1,2)$ and see that they tend to accumulate in specific values. We then find that the order distance of two $\pi_\beta^n,\pi_\beta^m$ that are initially (for $\beta=2$) close tends to grow exponentially in $\beta$ as $\beta$ decreases to 1. We also show how the ordering in $\Omega_{\beta,N}$ changes altogether by plotting the set of all points $\{n,\pi_\beta^n\}$ for different values of $\beta$, where $n$ is the lexicographical ordering. Finally, we introduce a finite analog to the $\beta$-compactum and show that the order plot of $\Omega_{1/\beta,N}$ of when $\beta$ is the golden ratio $g=(1+\sqrt5)/2$ resembles the Binary Reflected Gray Code (BRGC).

\section{Physical Motivation}

There have been some parallels between Bernoulli convolutions and physics. The first example is the Fat Baker's Map, which is a generalization of the Baker's map\cite{Alexander_Yorke_1984}, a classical example of a chaotic dynamical system. The fat baker's map $T_\lambda: A\rightarrow A$ is a map from the square $A=\{(x,y):-1\leq x \leq1,-1\leq y \leq1\}$ to itself with $0<\lambda<1$ such that

\begin{equation}
\label{eq:Fat baker's map}
T_\lambda(x,y)=  \begin{cases}
 \{(\lambda x+(1-\lambda),2y-1\} & \ y \geq 0 \\
 \{(\lambda x-(1-\lambda),2y+1\} & \ y < 0
  \end{cases}
\end{equation}

The regular Baker's map is recovered when $\lambda=1/2$. The Sinai-Ruelle-Bowen measure of $T_\lambda$ is then $\nu_\lambda$ times a uniform distribution in the y axis.\par
There have also been works relating Bernoulli convolutions to random walks in 1D, where each step decreases by a factor of $\lambda$, see e.g. \cite{PhysRevE.73.036118,10.1119/1.1632487,PhysRevE.62.7748}. Lastly, we point to novel possible connection to physics - relating to the problem of intermittency of the energy cascading in turbulent flows. The velocity fields of fluids are governed by the Navier-Stokes equations, and for high Reynolds numbers they are known to be chaotic. Even so, there is extra structure in the statistics of the velocity fields, which relates to the cascading of kinetic energy from the large scale $l_0$ where the energy is pumped into the system, to the small scale $\eta$ where the energy dissipates to heat through viscosity. In between these scales is a range called the 'inertial range' $\eta\ll l \ll l_0$, where viscous forces are negligible, and a universal pattern emerges, seen through the structure functions $S_n(l)=\langle |\delta u_l|^p\rangle\sim l^{\zeta_p}$ and $\langle \epsilon_l^p\rangle\sim l^{\tau_p}$, where $|\delta u_l|$ is the difference at length $l$ of the velocity field, and $\epsilon_l$ is the coarse-grained energy dissipation fluctuation from the mean value $\bar{\epsilon}$ with $\epsilon\sim(du/dt)^2$, averaged over a ball of radius $l$. The two scaling exponents are related by the Kolmogorov Refined Similarity Hypothesis  

\begin{equation}
\label{eq:KRSH}
\zeta_p=\frac{p}{3}+\tau_{p/3}
\end{equation}

and it has been an effort over the years to find toy models that reproduce measures that obey these moment structures. Further restrictions motivated by experiments, namely that the energy dissipation field, which is observed to be highly intermittent and has its intense high values concentrated at sparse, scale dependent regions of the flow, call for the coarse grained dissipation field to be a multifractal measure. See \cite{FrischSulemNelkin1978,BenziPaladinParisiVulpiani1984,MeneveauSreenivasan1987,Castaing1989,AndrewsPhillipsShivamoggiBeckJoshi1989,Yamazaki1990,PhysRevLett.72.336} for a non-exhaustive list of examples of such models. Out of these, it is worth mentioning the $p$-model of \cite{MeneveauSreenivasan1987} for its similarity to the IFS approach to the finite Bernoulli convolutions measure - this model assumes that at step $k$, a $2^{k-1}$ region from the original space is split into two, with an equal probability of obtaining a value $p$ in one half, and a value of $1-p$ in the other. The resulting measure, when properly normalized, is multifractal with a singularity spectrum $f(\alpha(x))=-x\log_2(x)-(1-x)\log_2(1-x)$ and $\alpha=-x\log_2(p)-(1-x)\log_2(1-p)$ where $0\leq x \leq 1$ is the ratio of the number of steps with weight $p$ and the total number of steps. Most of these toy models have a singularity spectrum that is smooth in $\alpha$ and also concave, and so can relate to the $L^p$-spectrum $\tau_p$ by a Legendre transformation\footnote{In physics one often uses the generalized dimension $D_p=(1-p)\tau_p$\cite{PhysRevA.33.1141}}, but some recent phenomenological and experimental studies of the energy dissipation cascade emphasize rare intense events such as vortex tubes\cite{PhysRevE.67.016305} and shear layers\cite{10.1098/rspa.2020.0591} as the main dissipation mechanisms \cite{fractalfract6100613}. To this extent, the Bernoulli convolutions measure $\nu_\lambda(x)$ can serve as an interesting toy model for the energy dissipation field $\epsilon(x)$: for $1/2<\lambda<1$, Jordan, Shmerkin and Solomyak showed in \cite{JORDAN_SHMERKIN_SOLOMYAK_2011} that for Lebesgue-a.e. $\lambda\in(1/2,1)$, the measure $\nu_\lambda$ is absolutely continuous and its local dimension satisfies $d_{\nu_\lambda}(x)=1$ for $\nu_\lambda$-a.e. $x$, where 

\begin{equation}
\label{eq:local_singularity_dimesnion}
d_{\nu_\lambda}(x)=\lim_{r\rightarrow 0}\frac{\ln{\nu_\lambda \left(B(x,r)\right)}}{\ln{r}}
\end{equation}

and $B(x,r)$ being a ball of radius $r$ centered at $x$. At the exceptional set of points, where many cylinders overlap, the local dimension should look like

\begin{equation}
\label{eq:local_singularity_dimesnion_overlaps}
d_{\nu_\lambda}(x)\approx\frac{\ln{2-\lim_{n\rightarrow\infty}\frac{1}{n}\ln{N_n(x)}}}{\ln{1/\lambda}}
\end{equation}

Where $N_n(x)$ is the number of $n$-binary words that overlap in $x$, which will be defined in sec. \ref{crossings}, eq. \ref{inverse pi n}. \cite{JORDAN_SHMERKIN_SOLOMYAK_2011} also shows that even for $\nu_\lambda(x)$ that is absolutely continuous the local dimension level sets can be non-empty and even have positive Hausdorff dimension. It is then a possibility that for the right parameter $\lambda$, $\nu_\lambda(x)$ might be an adequate model for intermittent turbulence, which is 'weakly multifractal' and has a singularity spectrum that fits experiments.

\section{Crossings}\label{crossings}

To clarify the connection to $\beta$-expansions, from now on we will use $\beta=1/\lambda$. In the case of the full (one-sided) shift $\mathcal{A}^\mathbb{N}$, the set of sequences that are coefficients of a $\beta$-expansion of $x$, defined as $\mathcal{E}_\beta(x):=\pi_\beta^{-1}(x)$, is infinite with a continuum  cardinality for Lebesgue a.e. $x\in I_\beta=[0,\beta/(\beta-1))$ \cite{Sidorov1}.\par

It is useful to see how this continuum emerges by looking at a finite version of $\mathcal{E}_\beta(x)$ 

\begin{equation}
\label{inverse pi n}
\mathcal{E}^N_\beta(x):=\{(\epsilon_0,\epsilon_1,\dots,\epsilon_{N-1})\in\mathcal{A}^N|\exists(\epsilon_{N},\epsilon_{N+1},\dots):x=\sum_k \epsilon_k \beta^{-k}\}
\end{equation}

which is the set of all $N$ words that have a tail such that the whole sequence is a $\beta$-expansion of $x$.
It was shown in \cite{kempton2012} that the size $\mathcal{N}^N_\beta(x):=|\mathcal{E}^N_\beta(x)|$
for a.e. $\beta\in(1,2)$, a.e. $x\in I_\beta$ grows like

\begin{equation}
\label{growth rate general}
\lim_{N\rightarrow\infty}\sup\frac{1}{N}\log{\mathcal{N}^N_\beta(x)}=\log\left( \frac{2}{\beta}\right)
\end{equation}

and that (\ref{growth rate general}) is strictly less than $\log\left(2/\beta\right)$ when $\beta$ is a Pisot number \cite{feng2009growthratebetaexpansions}.\par

We suggest a different approach- begin with the finite set of binary words $\mathcal{A}^N$, and the set of their $\beta$-expansions $\Omega_{\beta,N}$. A Plot of $\Omega_{\beta,N}$ for $\beta\in(1,2)$ and $N=9$ is seen in fig. (\ref{Normalized Polynomials n=9}), where it is clear that the crossings are distributed in $\beta$ in a very non-uniform manner, mostly accumulating in specific values of $\beta$, seen in fig. (\ref{Normalized Polynomials n=9}) as 'bright columns'. Since the crossings are values of $x$ and $\beta$ for which two or more $\beta$-expansions coincide, by defining the difference of each pair of $\beta$-expansions as the polynomials $$p_\beta^{n,m}:=\pi_\beta^n-\pi_\beta^m=\sum_k \xi^{n,m}_k \beta^{-k}$$ the crossings become the real roots of $p_\beta^{n,m}$.\par We call $P_{\beta,N}$  the (multi)set of all $p_\beta^{n,m}$. Since every sequence $\epsilon_n$ is made of 0 and 1, the coefficients of $\xi^{n,m}_k$ are drawn from $\{-1,0,1\}$ \footnote{This is different from the symmetric Bernoulli convolution where the coefficients are $\{-1,1\}$, which leads to $p_\beta^{n,m}$ having coefficients $\{-2,0,2\}$, but since we're interested in the roots of $P_{\beta,N}$ this overall factor of 2 makes no difference}. Moreover, $\Omega_{\beta,N}$ consists of the $\beta$-expansions of all possible binary sequences of length $N$, so all possible polynomials of degree $N$ with coefficients $\{-1,0,1\}$ are in $P_{\beta,N}$. The connection between Bernoulli convolutions and polynomials with  $\{-1,0,1\}$ coefficients is well known, e.g. Hochman \cite{Hochman_2014} has shown that a dimension drop in the measure $\nu_\lambda$ is related to the roots (or sufficiently small size) of such polynomials for the parameter $\lambda$ by exact (or nearly exact) overlaps of contractions in IFS.\par
The degeneracies of $p_\beta^{n,m}\in P_{\beta,N}$ correspond to the values of $0$ in $\xi^{n,m}$, which can come from $(\epsilon^n)_k$ and $(\epsilon^m)_k$ having both $1$ or $0$ as the $k$ digit, so one component in the degeneracy of $p_\beta^{n,m}$ is $2^{\alpha(\xi^{n,m})}$, where $\alpha$ counts the number of zeros in the coefficients $\xi^{n,m}$.\par
\begin{figure}[!htbp]
    \centering
    \includegraphics[width=0.65\linewidth]{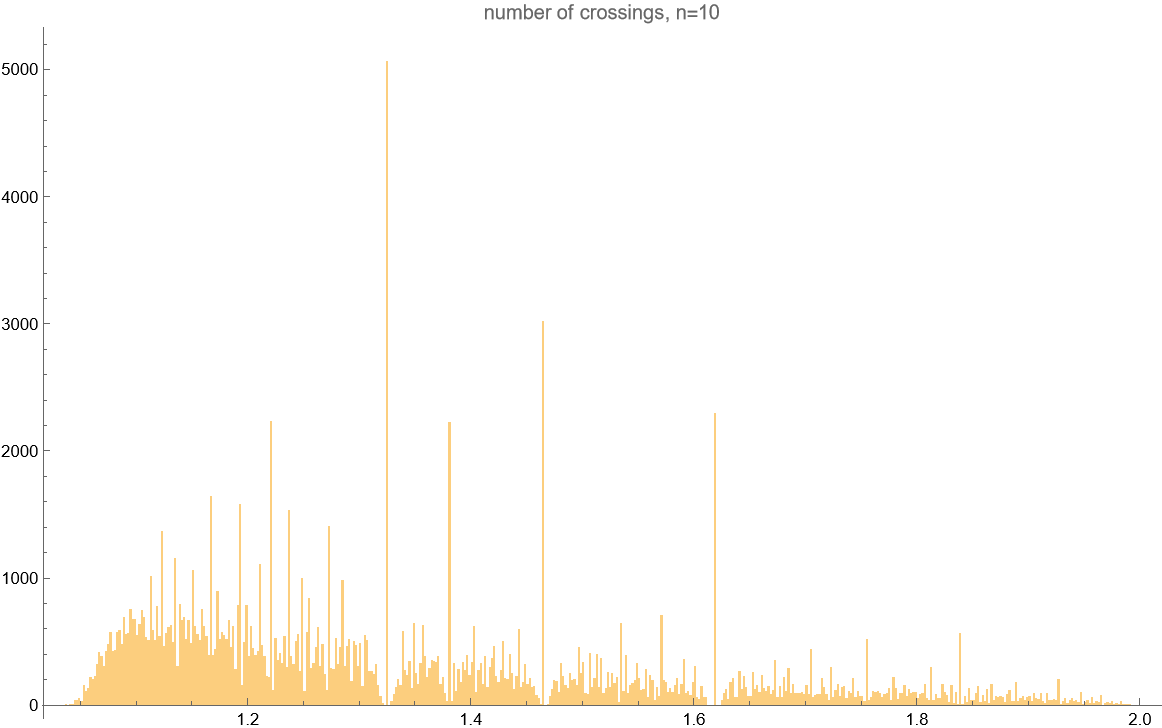}
    \caption{Distribution of the number of real roots in the interval $(1,2)$ of the elements of $P_{\beta,N}$ with $N=10$. Some of the noticeable peaks are Pisot numbers, but many (especially the cases with $\beta<\beta_2$, which is the smallest Pisot number) are not. The highest peaks map to the bright columns in fig. (\ref{Normalized Polynomials n=9}), and we can see that the "clearing" of crossings in the surrounding of the peaks is the reason the columns seem to be bright.} 
    \label{Number of crossings n=10}
\end{figure}

By numerical calculation, we get the distribution of the crossings in $\beta$ for $N=10$, seen in Fig. (\ref{Number of crossings n=10}), which confirms the picture from fig.(\ref{Normalized Polynomials n=9}), where the bright columns correspond to values of $\beta$ where the number of crossings is much higher than the surrounding values.

Out of these, some are Pisot numbers, but all values smaller than $\beta_2\approx1.32472$ (also called the Plastic constant, which is the smallest Pisot number) are not. Most prominent of these for $N=10$ are $\beta\approx1.22074$ and $\beta\approx1.23651$ which are the largest real roots of $x^4-x-1=0$ and $x^5-x^3-1=0$ respectively.
We suggest that calculations of the growth rate of the number of crossings for specific values of $\beta$ can show which will remain dominant when $N\rightarrow\infty$. One can also use the relatively small value of $N$ to directly calculate the discrete entropy of (\ref{eq:finite BC}), where the entropy is the information entropy, defined for a measure $\mu$ on a discrete set $\{x_1,x_2\dots\}$ as

\begin{equation}
\label{information entropy}
H(\mu):=-\sum_i\mu(x_i)\log\left(\mu(x_i)\right) 
\end{equation}

The Garsia entropy\cite{Garsia63} $h_\lambda$ is then

\begin{equation}
\label{Garsia entropy}
h_\lambda:=\lim_{N\rightarrow\infty}\frac{H\left(\nu_{\lambda,N}\right)}{N} 
\end{equation}

Hochman's entropy formula\cite{Hochman_2014} implies that if $h_\lambda<\log\lambda^{-1}$, then $\dim_H\nu_\lambda<1$ and thus $\nu_\lambda$ is non-trivially singular continuous. Calculating $H\left(\nu_{\lambda,N}\right)/N$ directly for $\beta$ with a large number of crossings, especially for the non-Pisot values, might then be used to extrapolate $h_\lambda$ and indicate numerically if they are in the exceptional set.

\section{Order dynamics}\label{Order dynamics}

The order of a $\beta$-expansion $g_n:=g(\pi_\beta^n)$ (\ref{eq:order of pi}), is a fairly erratic function of $\beta$. One can see that picking a random $\beta$-expansion in \ref{Normalized Polynomials n=9} when $\beta=2$ and tracing its trajectory by eye as $\beta$ decreases becomes a difficult task when $\beta\lesssim1.7$ because of the numerous crossings. Each crossing of two $\pi$'s switches their order. If a crossing is between more than two $\pi$'s, as is mostly the case as $N$ gets larger, the change in the order gets more complicated and depends on the derivatives of the expansions at the crossing value $\beta$. This also means that the order of all $\pi$ in a single crossing is the same, and that order value of $\pi_\beta^n$ for a given $\beta$ increases discretely in jumps the size of the number of $\pi$'s per crossing.\par
We can define the'order distance' between two expansions for a given $\beta$

\begin{equation}
\label{eq:order distance}
\gamma(\pi^n_\beta,\pi^m_\beta):=|g(\pi^n_\beta)-g(\pi^m_\beta)|
\end{equation}

When $N\rightarrow\infty$ for a.e. $\beta\in(1,2)$, this definition should weak-* converge to

\begin{equation}
\label{eq:order distance integral}
\gamma(\pi^n_\beta,\pi^m_\beta)=\left|\int_{\pi_\beta^m}^{\pi_\beta^n}d\nu_{1/\beta}\right| 
\end{equation}

\begin{figure}[htbp]
    \centering
    \begin{subfigure}[h]{0.47\textwidth}
         \centering
    \includegraphics[width=\linewidth]{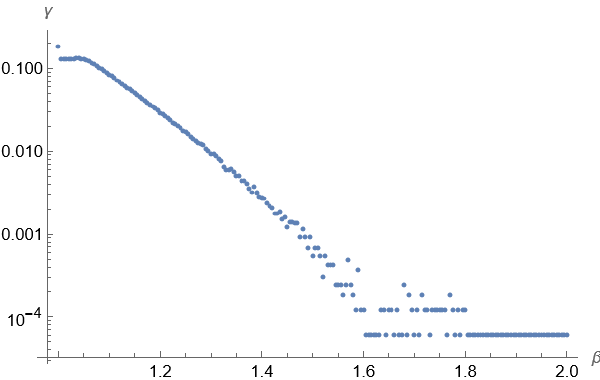}
    \caption{} 
         \label{fig:Ordering distance Log-Log}
     \end{subfigure}
     \hfill
    \begin{subfigure}[h]{0.47\textwidth}
         \centering
         \includegraphics[width=\textwidth]{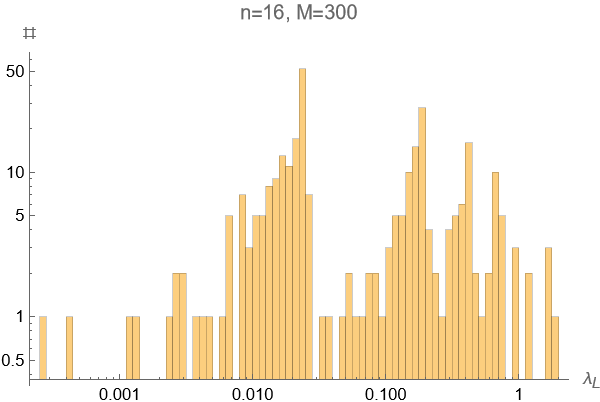}
         \caption{}
         \label{fig:Lyapunov Log-Log distribution}
     \end{subfigure}
     \hfill    
         \caption{Fig. (\ref{fig:Ordering distance Log-Log}) shows a Log-Linear plot of a numerical calculation of the ordering distance $\gamma$ between the $\beta$-expansions of two lexicographically adjacent sequences. The exponential growth can be seen as a linear slope. The Plateau at small values of $\beta$ is a consequence of the lack of crossings in the region, which is due to the finiteness of the system. This effect is also seen in fig. (\ref{Normalized Polynomials n=9}) and (\ref{Number of crossings n=10}). In fig. (\ref{fig:Lyapunov Log-Log distribution}) we see a histogram of 300 Lyapunov exponents calculated for different sequences which are equally distributed in the whole $\Omega_{\beta,N}$ with $N=16$. The plot shows a tendency towards 3 different clusters of exponents, for each we've calculated the means.} 
        \label{Ordering distance}
\end{figure}

We test the mixing in the ordering of $\beta$-expansions as $\beta$ changes numerically by fixing a finite $N$, choosing two lexicographically adjacent sequences $\epsilon_n,\epsilon_{n+1}$, and computing their order distance $\gamma(\pi^n_\beta,\pi^{n+1}_\beta)$ for small changes in $\beta$. Beginning at $\beta=2$, $\gamma=2^{-N}$  for all choices of sequences since every $\beta$-expansion is the unique binary expansion and is at equal distance from its adjacent expansions.
The typical result shows an exponential growth of $\gamma$ as $\beta$ decreases from 2 to 1, as seen in figure (\ref{Ordering distance}). From numerical calculations, we find that the Lyapunov exponent has a distribution with three main clusters, each having a mean value of $\lambda_{L1}\approx0.0175,~ \lambda_{L2}\approx0.1557,~ \lambda_{L3}\approx0.3861$. See (\ref{Lyapunov exponents}) for more details.

Note that the change in the order distance $\gamma(\pi^n,\pi^m)$ depends on the change of order of the boundaries $g(\pi^i)$ as $\beta$ decreases, the latter are local values that depend only on the number of other $\pi$'s the specific $\pi^i$ crosses on that specific $\beta$. Heuristically, since the total distance increases, these crossing must have a net negative 'flux' of trajectories, meaning that the number of trajectories that cross into the interval from both ends is larger than the number of trajectories that cross out.\par
\begin{figure}[htbp]
 \centering
     \begin{subfigure}{0.49\textwidth}
         \centering
         \includegraphics[width=\textwidth]{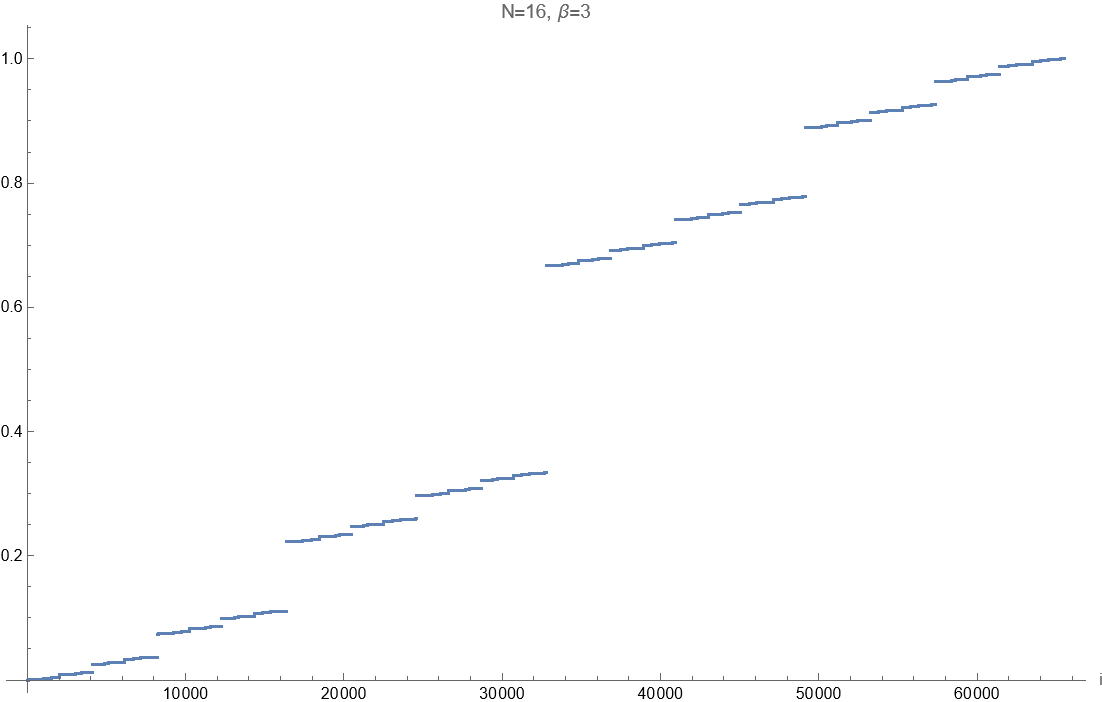}
         \caption{$\beta=3$}
         \label{fig:Order plot a=3}
     \end{subfigure}
     \hfill
     \begin{subfigure}{0.49\textwidth}
         \centering
         \includegraphics[width=\textwidth]{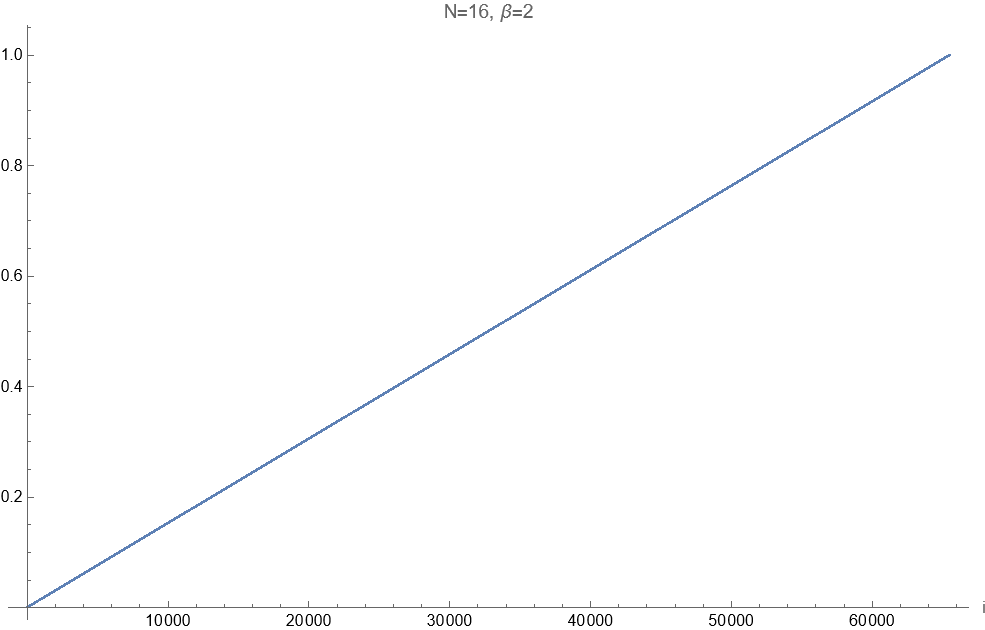}
         \caption{$\beta=2$}
         \label{fig:Order plot a=2}
     \end{subfigure}
     \hfill
      \begin{subfigure}{0.49\textwidth}
         \centering
         \includegraphics[width=\textwidth]{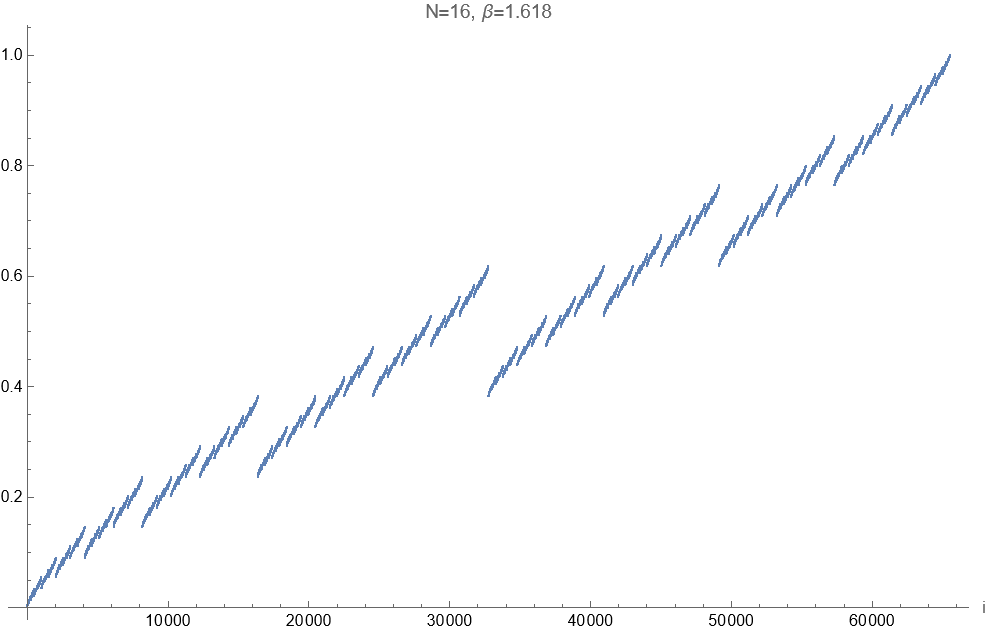}
         \caption{$\beta=1.618$}
         \label{fig:Order plot a=Phi}
     \end{subfigure}
     \hfill
           \begin{subfigure}{0.49\textwidth}
         \centering
         \includegraphics[width=\textwidth]{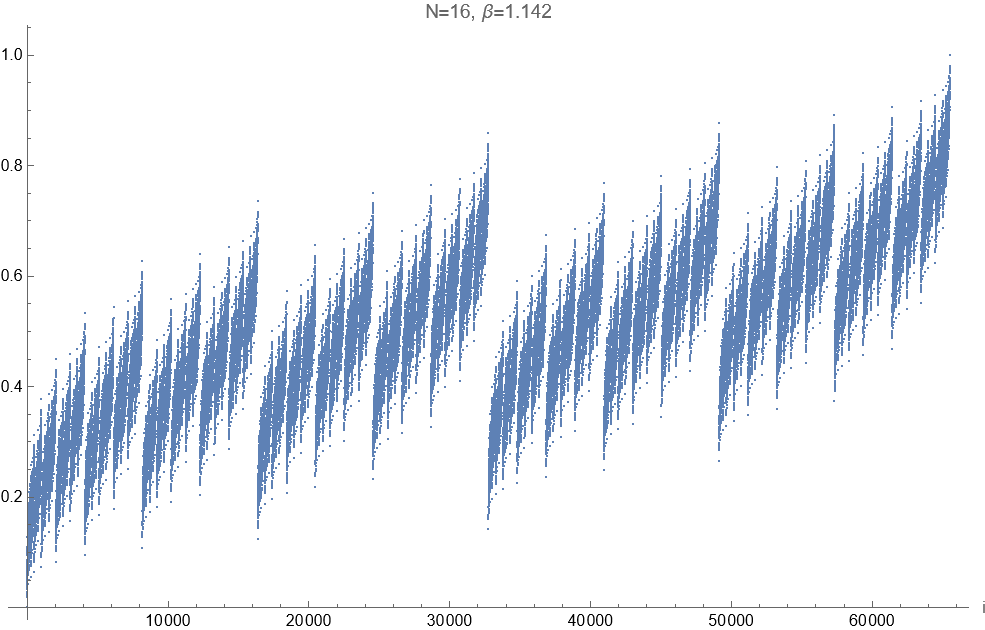}
         \caption{$\beta=1.142$}
         \label{fig:Order plot a=1.1.142}
     \end{subfigure}
     \hfill
           \begin{subfigure}{0.49\textwidth}
         \centering
         \includegraphics[width=\textwidth]{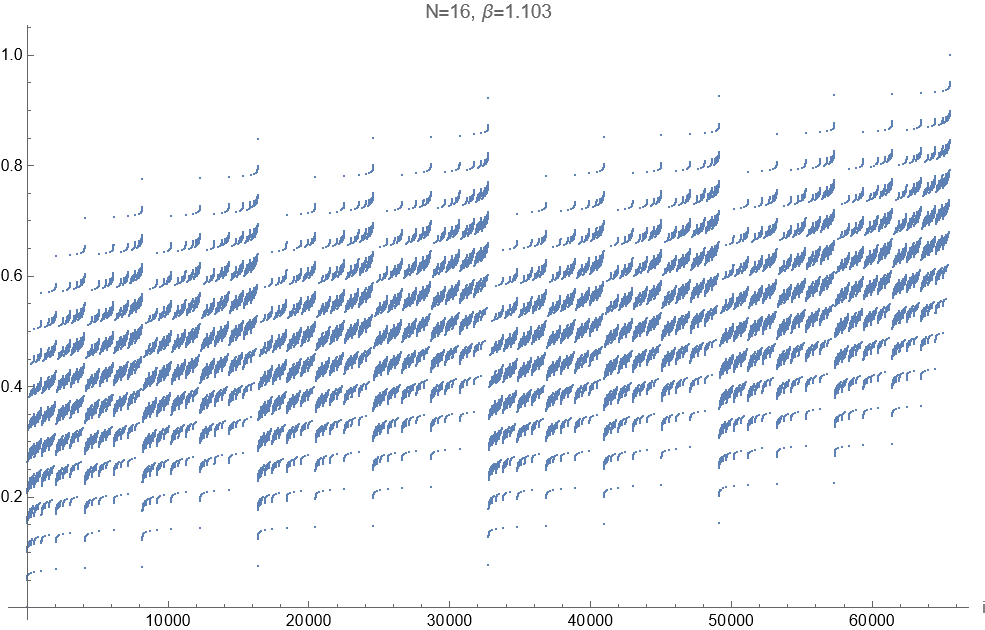}
         \caption{$\beta=1.03$}
         \label{fig:Order plot a=1.03}
     \end{subfigure}
     \hfill
        \begin{subfigure}{0.49\textwidth}
         \centering
         \includegraphics[width=\textwidth]{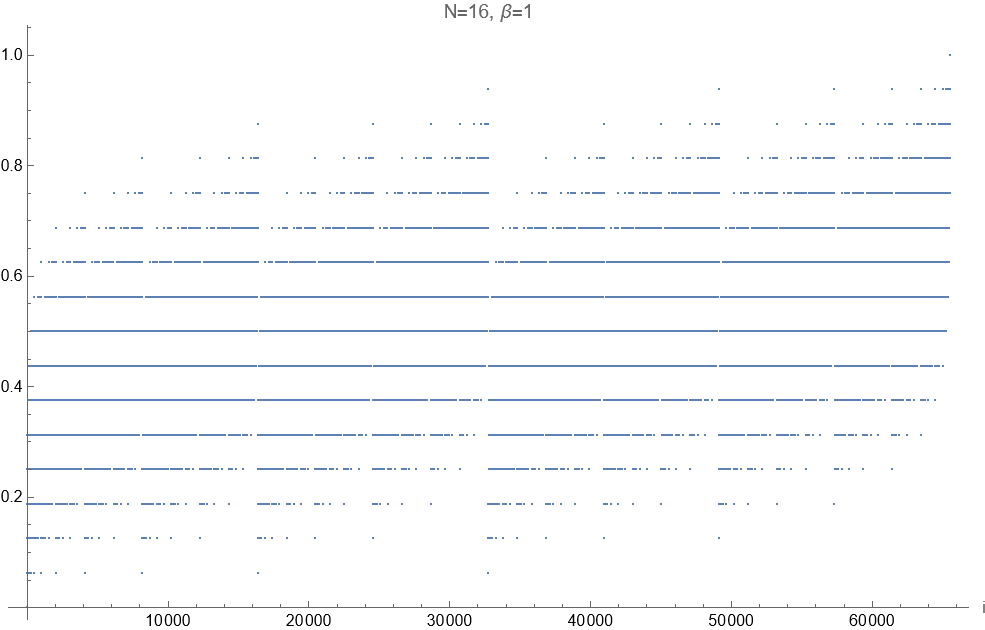}
         \caption{$\beta=1$}
         \label{fig:Order plot a=1}
     \end{subfigure}
     \hfill
    \caption{Order plots of $\Omega_{\beta,N}$ for $N=16$ and various $\beta>1$, where the order is given by the lexicographical order of the sequences in the $\beta$-expansions.}
    \label{fig:Order plots beta>1}
\end{figure}

\begin{figure}[htbp]
 \centering
     \begin{subfigure}{0.49\textwidth}
         \centering
         \includegraphics[width=\textwidth]{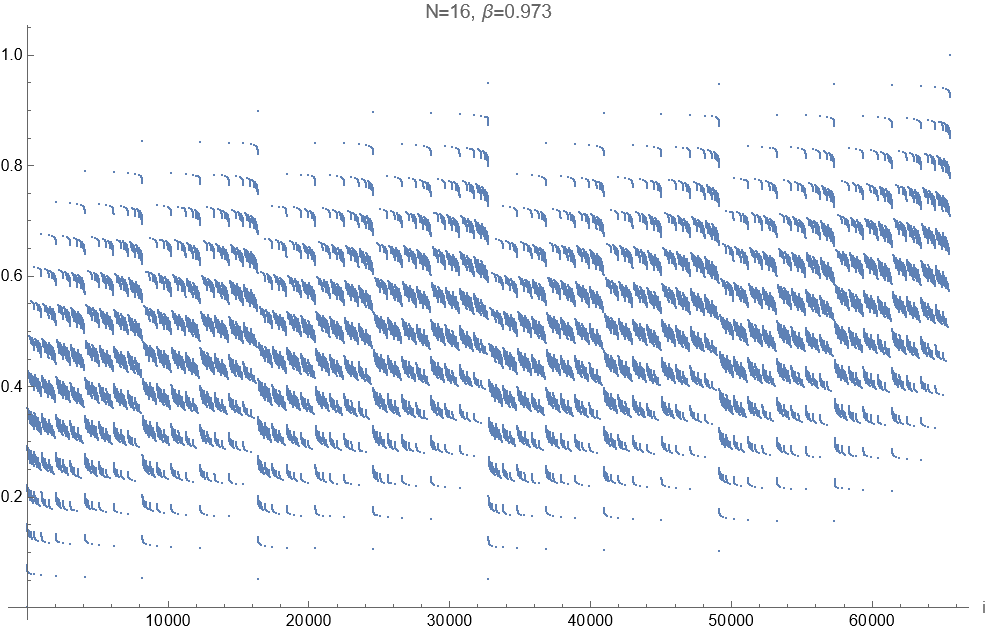}
         \caption{$\beta=0.973$}
         \label{fig:Order plot a=0.973}
     \end{subfigure}
     \hfill
     \begin{subfigure}{0.49\textwidth}
         \centering
         \includegraphics[width=\textwidth]{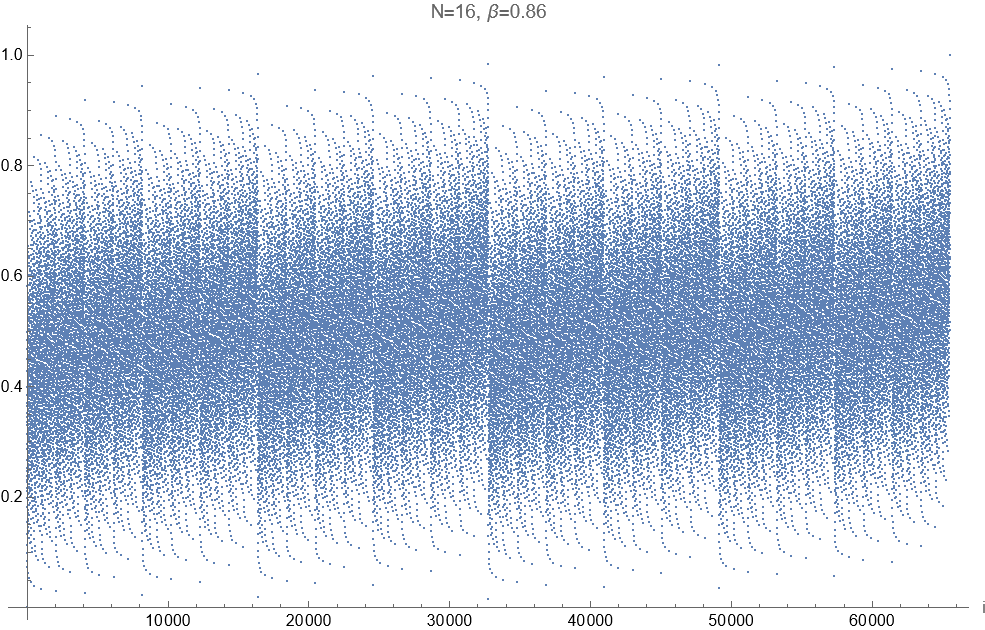}
         \caption{$\beta=0.86$}
         \label{fig:Order plot a=0.86}
     \end{subfigure}
     \hfill
      \begin{subfigure}{0.49\textwidth}
         \centering
         \includegraphics[width=\textwidth]{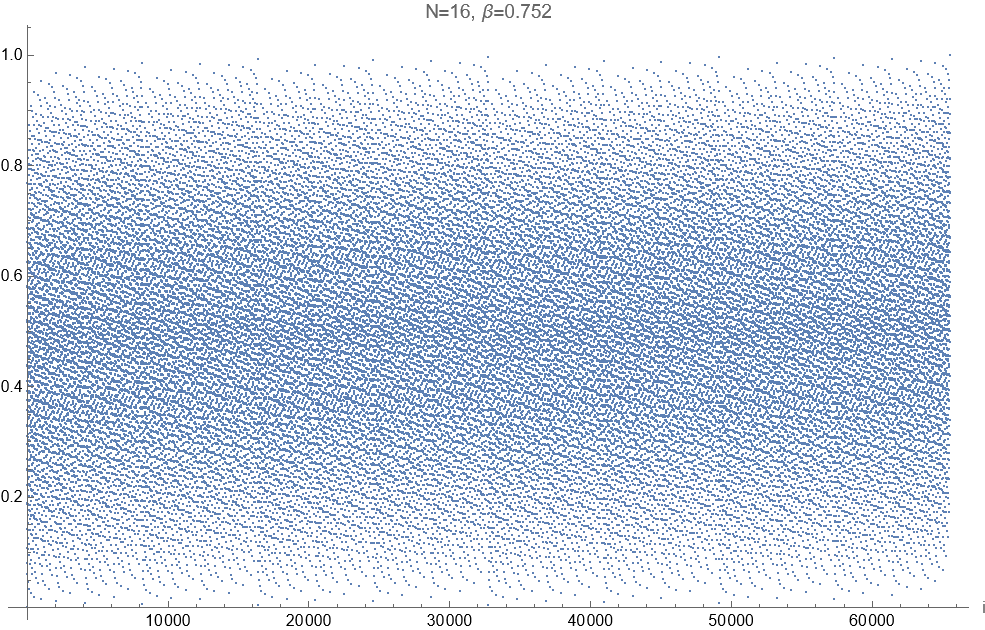}
         \caption{$\beta=0.752$}
         \label{fig:Order plot a=0.752}
     \end{subfigure}
     \hfill
           \begin{subfigure}{0.49\textwidth}
         \centering
         \includegraphics[width=\textwidth]{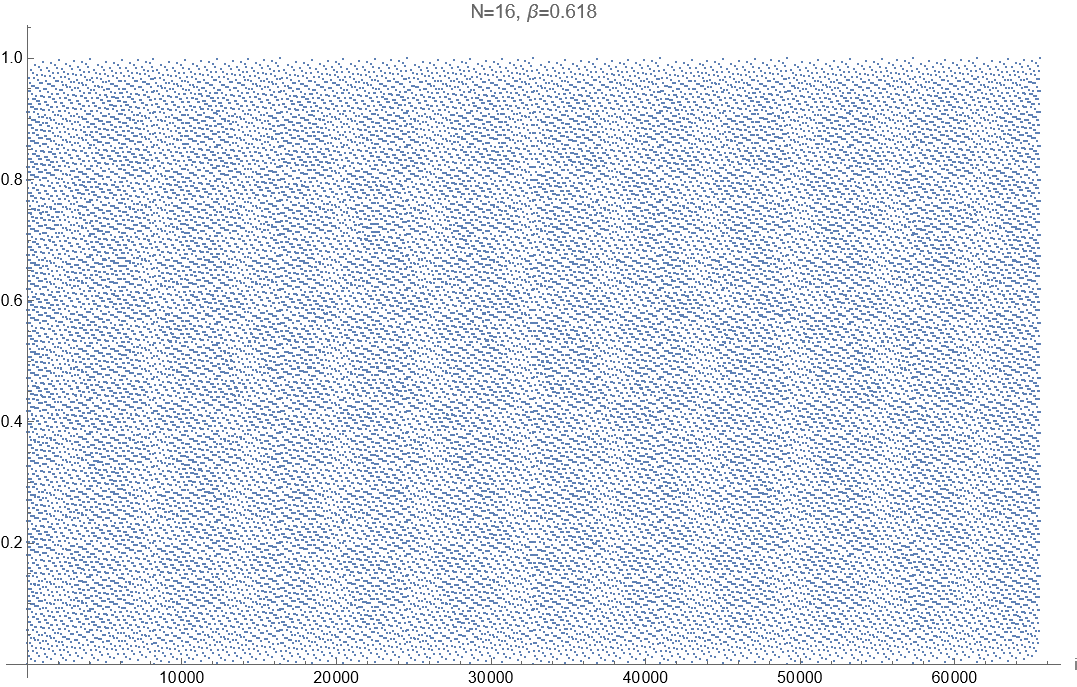}
         \caption{$\beta=0.618$}
         \label{fig:Order plot a=0.618}
     \end{subfigure}
     \hfill
           \begin{subfigure}{0.49\textwidth}
         \centering
         \includegraphics[width=\textwidth]{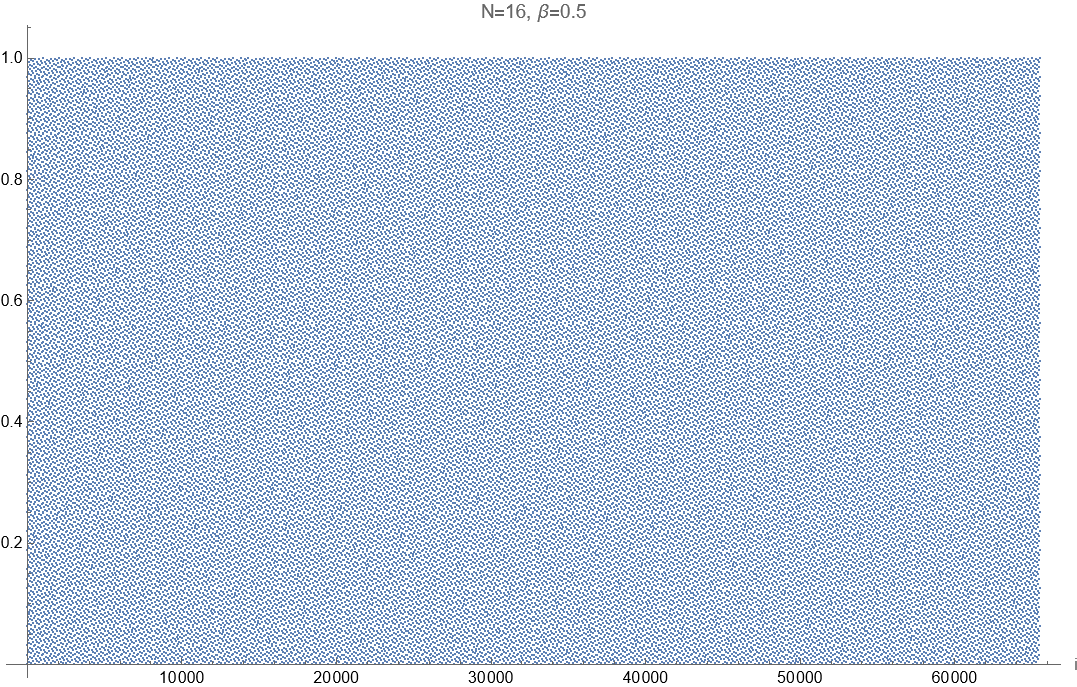}
         \caption{$\beta=0.5$}
         \label{fig:Order plot a=0.5}
     \end{subfigure}
         \hfill
         \begin{subfigure}{0.49\textwidth}
         \centering
         \includegraphics[width=\textwidth]{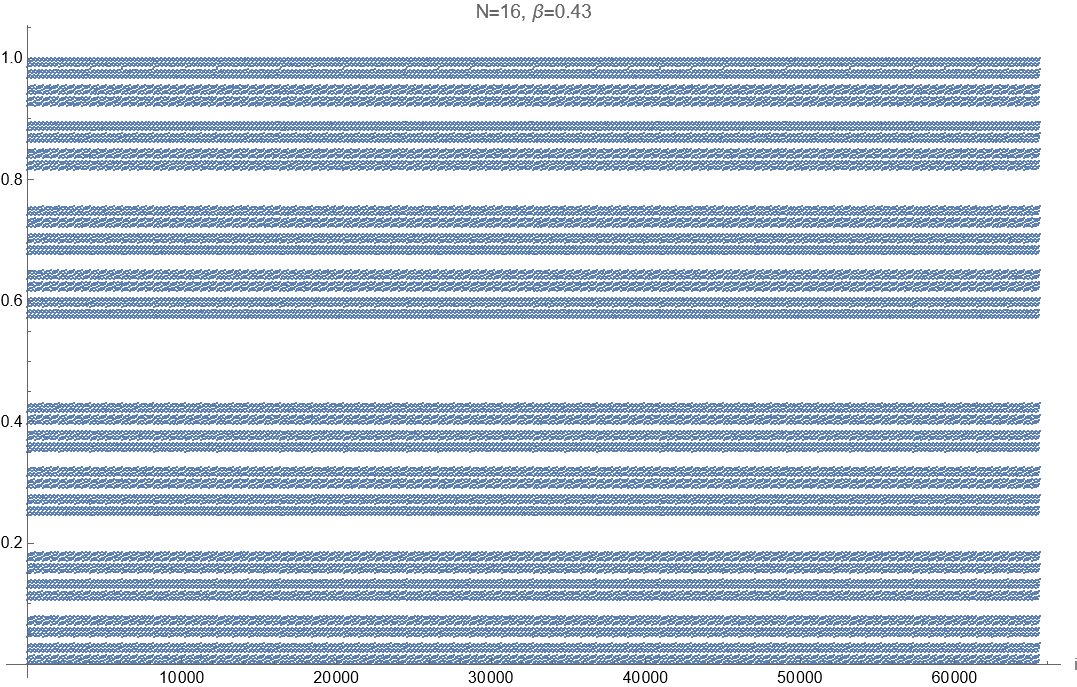}
         \caption{$\beta=0.43$}
         \label{fig:Order plot a=0.43}
     \end{subfigure}
     \hfill
    \caption{Order plots of $\Omega_{\beta,N}$ for $N=16$ and various $\beta<1$. In some plots, such as fig. (\ref{fig:Order plot a=0.752}) and (\ref{fig:Order plot a=0.618}), horizontal and diagonal patterns are visible. These indicate correlations between projections that are lexicographically distant.}
    \label{fig:Order plots beta<1}
\end{figure}

 
We can also visualize the order of $\Omega_{\beta,N}$ and the way that order changes with $\beta$ by plotting pairs of $\{n,\pi_\beta^n\}$ for all $\pi^n_\beta\in\Omega_{\beta,N}$ and a fixed $\beta$. These plots can be seen in figure (\ref{fig:Order plots beta>1}) for different values of $\beta$. For example, (\ref{fig:Order plot a=2}) is the order plot for $\beta=2$, where the values of $\pi$ increases as $n$ increases. For smaller $\beta$ the plot fractures and each $\pi$ shifts in a different rate. When $\beta=1$ these shifts overlap to form the equal weighted binomial distribution. One can go further and take values of $\beta<1$, as seen in figure (\ref{fig:Order plots beta<1}). The projection onto the $y$ axis is the same for $\beta$ and $1/\beta$ (up to  a scaling factor) since $\Omega_{\beta,N}=\beta^{-(N-1)}\Omega_{1/\beta,N}$, even though the order plots look quite different due to the reversing of the digits. In particular, for $\beta=1/2$  the equidistribution of $\Omega_{\beta,N}$ is recovered, and for $\beta<1/2$ we get the finite versions of middle-$\alpha$ Cantor sets. The order plot of $\beta=1/2$ is of particular interest, since it seems not only to be marginally equidistibuted in both directions, but also equidistibuted in the whole order plane. Upon further inspection, this seems to rise from a quasi-periodic pattern that depends on $n$.

Following these order plots, we also looked at cases where $\Omega_{\beta,N}$ contains the $\beta$-expansions of a 'finite $\beta$-shift' (or $\beta$-compactum), which in the infinite case is made of all the inverse $\beta$-expansions of $x\in I_\beta$ by specifically choosing the greedy expansion. Alternatively, the $\beta$-shift is also the sub-shift space of all the shifts of the infinite sequence that is the greedy $\beta$-expansion of $x=1$ for a specific $\beta$ that are lexicographically smaller than it. Notice that the $\beta$-shift is $\{0,1\}^{\mathbb{N}}$ when $\beta=2$.\par
\begin{figure}[htbp]
    \centering
    \includegraphics[width=0.65\linewidth]{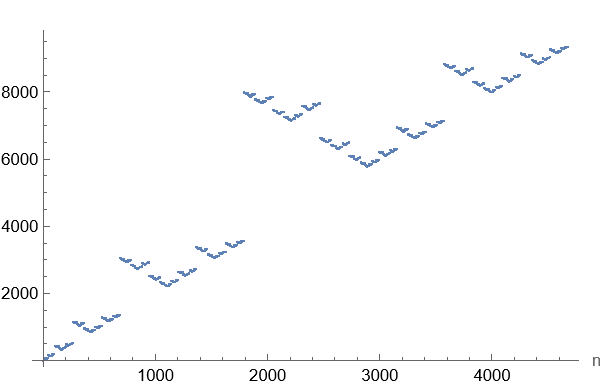}
    \caption{Order plot of $\Omega_{\beta,N}$ with sequences taken from the $\beta$-shift when $\beta$ is the golden ratio, plotted for $1/\beta$.}
    \label{Order plot 1/GR}
\end{figure}

For the finite case, we fixed a $\beta$ and chose the greedy expansion for a sequence of equally distributed $x_i\in I_\beta$. The order plot of $\Omega_{\beta,N}$ for this $\beta$-compactum is, by construction, a linear plot, like in fig. (\ref{fig:Order plot a=2}). We then plotted the order plot for $1/\beta$. For most $\beta$ this was not very insightful, having the expansions of the lexicographically smaller values becoming very large and the lexicographically larger values become small, due to the reflection of the order of the digits in the sequences when $\beta\rightarrow1/\beta$. Surprisingly, it turns out that for $\beta=g$ ($g$ being the  golden mean), this inversion of the order plot gives a specific fractal pattern seen in (\ref{Order plot 1/GR}).

\begin{figure}[htbp]
    \centering
    \includegraphics[width=0.65\linewidth]{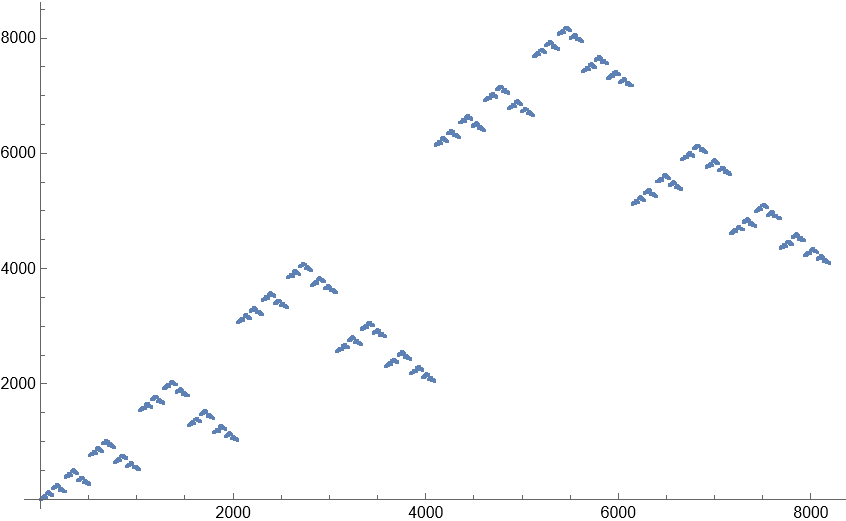}
    \caption{Order plot of the BRGC for strings of length $N=13$. Though similar to fig. (\ref{Order plot 1/GR}), the self - similar pattern seems inverted, and the projection onto the y-axis gives an equidistribution for all values, as opposed to fig. (\ref{Order plot 1/GR}) where the projection to the y-axis gives equidistributions but with gaps between the projections of the self-similar copies.}
    \label{BRGC}
\end{figure}

The pattern resembles (though in an inverted way and with gaps between the self similar patterns) the ordering of Binary Reflected Gray Code (BRGC), which is an alternative ordering of the binary numbers such that every consecutive pair of numbers differ by a change of only one digit (see \cite{mutze2024} for a contemporary review). The plot of the ordering of BRGC is seen in fig. (\ref{BRGC}). The connection between these two phenomena is an open question.  

\section{Acknowledgement}\label{Acknowledgement}

 We would like to thank Boris Solomyak, Uri Peleg, Yoav Zigdon, Lily Reisch - Kuritzky and Zamir Heller - Algazi for useful discussions, and especially to my advisor Cobi Sonnenschein for all the discussions and support along the way.
 This work  was supported in part by a grant 01034816 titled “String theory reloaded-
from fundamental questions to applications” of the “Planning and budgeting committee”.

\bibliographystyle{unsrt}
\bibliography{bibliography}

\begin{appendix}

\section{Lyapunov exponents}\label{Lyapunov exponents}

\begin{figure}[!htbp]
 \centering
     \begin{subfigure}{0.47\textwidth}
         \centering
         \includegraphics[width=\textwidth]{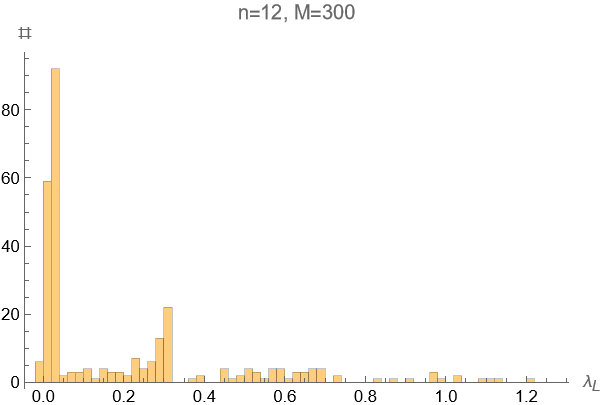}
         \caption{}
         \label{fig:ordering Lyapunov exponent, n=12, M=300}
     \end{subfigure}
     \hfill
     \begin{subfigure}{0.47\textwidth}
         \centering
         \includegraphics[width=\textwidth]{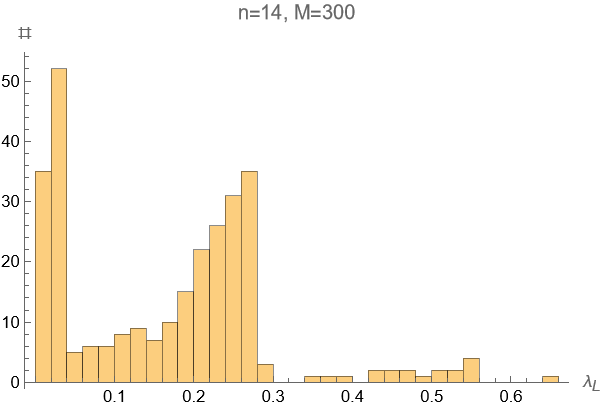}
         \caption{}
         \label{fig:ordering Lyapunov exponent, n=14, M=300}
     \end{subfigure}
     \hfill
      \begin{subfigure}{0.47\textwidth}
         \centering
         \includegraphics[width=\textwidth]{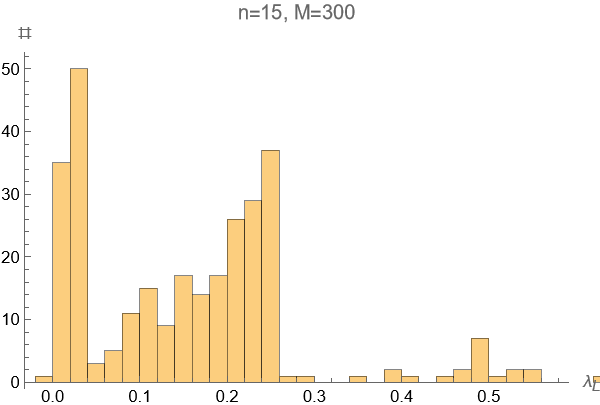}
         \caption{}
         \label{fig:ordering Lyapunov exponent, n=15, M=300}
     \end{subfigure}
     \hfill
        \begin{subfigure}{0.47\textwidth}
         \centering
         \includegraphics[width=\textwidth]{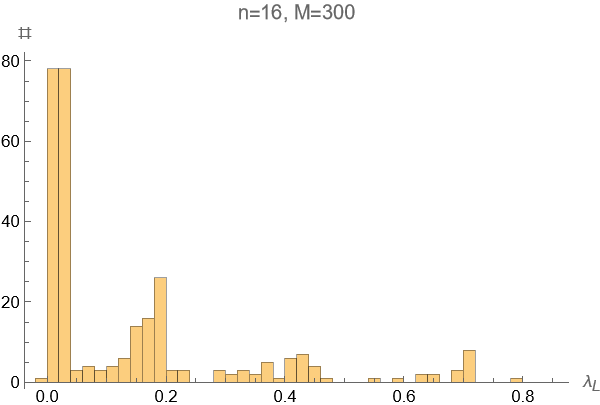}
         \caption{}
         \label{fig:ordering Lyapunov exponent, n=16, M=300}
     \end{subfigure}
     \hfill
    \caption{Histograms of Lyapunov exponents calculated for $M=300$ sequence pairs out of $\Omega_{\beta,N}$, with $N=12,14,15,16$. The clusters are visible for all values of $N$, though their relative sizes change. It is also worth noticing that the sharp edge of the second cluster, which is around $0.3$ when $N=12$, decreases as $N$ increases, and is around $0.2$ for $N=16$. Fig. (\ref{fig:Lyapunov Log-Log distribution}) is the Log-Linear plot of (\ref{fig:ordering Lyapunov exponent, n=16, M=300}).}
    \label{fig:ordering Lyapunov exponents M=300}
\end{figure}

The mean Lyapunov exponents in (\ref{Order dynamics}) were calculated by fixing some N, calculating $\Omega_{\beta,N}$, and choosing 300 pairs of lexicographically adjacent sequences that have an equal spacing in $\Omega_{\beta,N}$ when $\beta=2$. For each pair, the distance was calculated using (\ref{Ordering distance}) for 30 equally spaced values of $\beta$, between $\beta=1.4$ and $\beta=1.1$, from which the Lyapunov exponent was extracted by a linear fit of Log-Linear values. The mean is then calculated over these 300 Lyapunov exponents. We plot the histogram and Log-Log histogram of the Lyapunov exponents for $N=12,14,15,16$ in fig. (\ref{fig:ordering Lyapunov exponents M=300}). 

In Fig. (\ref{fig:Ordering distance Log-Log, first cluster}) we see a typical distance plot for a sequence pair which has a Lyapunov exponent in the first cluster ($\lambda_L<0.03$).

\begin{figure}[htbp]
    \centering
    \includegraphics[width=0.45\linewidth]{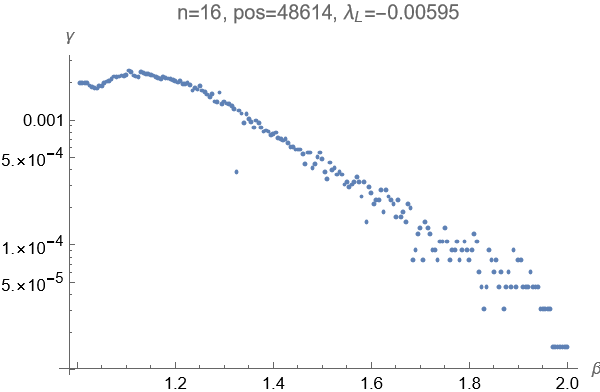}
    \caption{Plot of the ordering distance between two adjacent sequences when $N=16$, where the calculated Lyapunov exponent is in the first cluster (less than $0.03$). Notice that at small values of $\beta$ the distance stops increasing and shows a slight decrease.} 
    \label{fig:Ordering distance Log-Log, first cluster}
\end{figure}

\end{appendix}

\end{document}